\documentclass[prb,twocolumn,preprintnumbers,superscriptaddress,showpacs,nofootinbib]{revtex4}
\usepackage{amsmath,amssymb,bm,graphicx,hyperref}

\renewcommand\d{\partial}
\newcommand\+{\dagger}
\newcommand\x{{\bm{x}}}
\newcommand\p{{\bm{p}}}
\renewcommand\H{{\mathcal{H}}}
\newcommand\C{{\mathcal{C}}}
\newcommand\T{{\mathcal{T}}}
\newcommand\nr{\mathrm{nr}}

\begin{document}
\preprint{MIT-CTP 4165}

\title{Topological superconductors as nonrelativistic limits of \\
Jackiw-Rossi and Jackiw-Rebbi models}

\author{Yusuke Nishida}
\affiliation{Center for Theoretical Physics,
Massachusetts Institute of Technology, Cambridge, Massachusetts 02139, USA}
\author{Luiz Santos}
\affiliation{Physics Department, Harvard University, Cambridge, Massachusetts 02138, USA}
\author{Claudio Chamon}
\affiliation{Physics Department, Boston University, Boston, Massachusetts 02215, USA}

\begin{abstract}
 We argue that the nonrelativistic Hamiltonian of $p_x+ip_y$
 superconductor in two dimensions can be derived from the relativistic
 Jackiw-Rossi model by taking the limit of large Zeeman magnetic field
 and chemical potential.  In particular, the existence of a fermion zero
 mode bound to a vortex in the $p_x+ip_y$ superconductor can be
 understood as a remnant of that in the Jackiw-Rossi model.  In three
 dimensions, the nonrelativistic limit of the Jackiw-Rebbi model leads
 to a ``$p+is$'' superconductor in which spin-triplet $p$-wave and
 spin-singlet $s$-wave pairings coexist.  The resulting Hamiltonian
 supports a fermion zero mode when the pairing gaps form a hedgehoglike
 structure.  Our findings provide a unified view of fermion zero modes
 in relativistic (Dirac-type) and nonrelativistic (Schr\"odinger-type)
 superconductors.
\end{abstract}

\date{July 2010}

\pacs{74.90.+n, 74.20.Rp, 74.25.Ha, 03.65.Ge}

\maketitle

\section{Introduction}
Fermion zero modes bound to topological defects have been discovered by
Jackiw and Rebbi in 1976 (Ref.~\onlinecite{Jackiw:1975fn}) and recently
received renewed interest in condensed matter physics (see, for example,
Ref.~\onlinecite{Teo:2010zb}).  Vortices in a certain class of
superconductors in two dimensions (2D) support zero-energy Majorana
bound states and obey non-Abelian statistics~\cite{Read:1999fn}, which
can be potentially used for topological quantum
computation~\cite{Kitaev:1997wr}.  Although vortices in the ordinary
nonrelativistic $s$-wave superconductor do not support Majorana zero
modes, the weakly paired phase of the $p_x+ip_y$ superconductor, which
is believed to be realized in Sr$_2$RuO$_4$~\cite{Mackenzie:2003}, does
support Majorana zero modes bound to vortex
cores~\cite{Read:1999fn,Volovik:1999eh,Ivanov:2001}.

It is also known from the pioneering work by Jackiw and Rossi that the
relativistic $s$-wave superconductor in 2D (Jackiw-Rossi model) has
similar properties~\cite{Jackiw:1981ee}.  Remarkably it has been shown
that such a system can be realized on the surface of the
three-dimensional (3D) topological insulator in contact with the
$s$-wave superconductor~\cite{Fu:2008}.  Besides these examples, there
is a number of proposals to realize Majorana zero modes using
heterostructures of semiconductor and
superconductor~\cite{Sau:2010a,Alicea:2010,Sau:2010b}, superconductor
and ferromagnet~\cite{Lee:2009}, and quantum (anomalous) Hall state and
superconductor~\cite{Qi:2010}.

Although the nonrelativistic $p_x+ip_y$ superconductor and the
relativistic Jackiw-Rossi model share similar properties, the existence
of a fermion zero mode bound to a vortex has been discussed separately
in the two
systems~\cite{Jackiw:1981ee,Weinberg:1981eu,Tewari:2007,Gurarie:2007,Fu:2008,Fukui:2010a,Cheng:2010}.
In this paper (Sec.~\ref{sec:2D}), we argue that they are actually
linked by showing that the former Hamiltonian can be derived from the
latter by taking the limit of large Zeeman magnetic field and chemical
potential.  In particular, the fermion zero mode bound to a vortex
persists under taking this limit.

Then in Sec.~\ref{sec:3D}, we turn to the relativistic Jackiw-Rebbi
model in 3D, which is known to exhibit a fermion zero mode associated
with a pointlike topological defect
(hedgehog)~\cite{Jackiw:1975fn,Callias:1977kg,Teo:2009qv,Fukui:2010a}.
The limit of large mass and chemical potential (nonrelativistic limit)
leads to a ``$p+is$'' superconductor in which spin-triplet $p$-wave and
spin-singlet $s$-wave pairings coexist.  We show that the resulting
nonrelativistic Hamiltonian supports a fermion zero mode when the
pairing gaps form a hedgehoglike structure.

We note that the analysis presented in this paper is largely motivated
by the recent paper by Silaev and Volovik~\cite{Silaev:2010}:  The
nonrelativistic Hamiltonian of the Balian-Werthamer (BW) state of the
superfluid $^3$He was derived from the relativistic superconductor with
the odd parity pairing~\cite{Nishida:2010wr} and their topological
properties were studied.  In this paper, we shall broadly use
``relativistic'' to indicate Dirac-type Hamiltonians and
``nonrelativistic'' to indicate Schr\"odinger-type Hamiltonians.  For
readers' convenience, references to the main results are summarized in
Table~\ref{tab:summary}.

\begin{table}[b]
 \caption{References to the equations in which the Hamiltonian,
 zero-energy solution bound to a defect, and its normalizability
 condition are shown for the relativistic model and its nonrelativistic
 descendant both in 2D and 3D.  \label{tab:summary}}
 \begin{ruledtabular}
  \begin{tabular}{lccc}
   & Hamiltonian & Solution & Normalizability \\\hline
   2D relativistic & \eqref{eq:H_rela_2D}
       & \eqref{eq:solution_rela_2D} & \eqref{eq:condition_rela_2D} \\
   2D nonrelativistic & \eqref{eq:H_nonrela_2D}
       & \eqref{eq:solution_nonrela_2D} & \eqref{eq:condition_nonrela_2D} \\
   3D relativistic & \eqref{eq:H_rela_3D}
       & \eqref{eq:solution_rela_3D-1} or \eqref{eq:solution_rela_3D-2}
	   & \eqref{eq:condition_rela_3D} \\
   3D nonrelativistic & \eqref{eq:H_nonrela_3D}
       & \eqref{eq:solution_nonrela_3D-1} or \eqref{eq:solution_nonrela_3D-2}
	   & \eqref{eq:condition_nonrela_3D}
  \end{tabular}
 \end{ruledtabular}
\end{table}

\section{Jackiw-Rossi model in 2D and~its~nonrelativistic~limit \label{sec:2D}}

\subsection{Jackiw-Rossi model and fermion zero mode at a vortex}
We start with the Hamiltonian describing 2D Dirac fermions coupled with
an $s$-wave pairing gap (Jackiw-Rossi~\cite{Jackiw:1981ee} or
Fu-Kane~\cite{Fu:2008} model)
\begin{equation}
 H = \frac12\int\!d\x\,\Psi^\+\H\Psi
\end{equation}
with $\Psi^\+=(\psi^\+,-i\psi^T\sigma_2)$ and
\begin{equation}\label{eq:H_rela_2D}
 \H =
  \begin{pmatrix}
   \bm\sigma\cdot\p+\sigma_z h-\mu & \Delta \\
   \Delta^* & -\bm\sigma\cdot\p+\sigma_z h+\mu
  \end{pmatrix}.
\end{equation}
This Hamiltonian can be realized on the surface of the 3D topological
insulator in contact with the $s$-wave superconductor~\cite{Fu:2008}.
$h$ is the Zeeman magnetic field and $\mu$ is the chemical potential.
When the pairing gap $\Delta$ is spatially dependent,
$\p\equiv(p_x,p_y)$ has to be regarded as derivative operators
$(-i\d_x,-i\d_y)$.  The energy eigenvalue problem is
\begin{equation}\label{eq:eigenvalue_2D}
 \varepsilon
  \begin{pmatrix}
   u_1 \\ u_2 \\ v_2 \\ v_1
  \end{pmatrix}
  = \H
  \begin{pmatrix}
   u_1 \\ u_2 \\ v_2 \\ v_1
  \end{pmatrix}.
\end{equation}

When $h$ and $\mu$ are both zero, the number of fermion zero modes
($\varepsilon=0$) bound to a vortex formed by
$\Delta(x,y)\equiv\Delta_1+i\Delta_2$ is determined by the winding
number of the two scalar
fields~\cite{Jackiw:1981ee,Weinberg:1981eu,Fukui:2010a}
\begin{equation}\label{eq:index_2D}
 \mathrm{Index}\,\H = \frac1{2\pi}\int\!dl_i\,
  \epsilon_{ab}\hat\Delta_a\d_i\hat\Delta_b \equiv N_w,
\end{equation}
where $\hat\Delta_a\equiv\Delta_a/\sqrt{\Delta_1^2+\Delta_2^2}$ and the
line integral is taken at spatial infinity.  However, in the presence of
$h$ and $\mu$, the index theorem is no longer valid:  $h$ and $\mu$
terms in the Hamiltonian can couple zero modes and they become nonzero
energy states so that two states form a pair with opposite energies.
Therefore, in general, only one zero mode survives for odd $N_w$ while
no zero mode survives for even
$N_w$~\cite{Fukui:2010a,Teo:2010zb}.\footnote{Two exceptional cases are
  $h=\pm\mu$.  When $N_w>0\,(<0)$, the zero-energy solutions at
  $h=\mu=0$ still solve Eq.~(\ref{eq:eigenvalue_2D}) with
  $h=+(-)\mu\neq0$ and thus there are $|N_w|$ zero modes.  This can be
  easily seen if one rewrites the Hamiltonian in the basis where the
  chiral operator defined in Eq.~(\ref{eq:chiral_2D}) has the form
  $\chi=\mathrm{diag}(1,1,-1,-1)$ and recognizes that the $|N_w|$
  zero-energy solutions are eigenstates of $\chi$ with the eigenvalue
  $+1\,(-1)$ for $N_w>0\,(<0)$.}

If we work in polar coordinates $(r,\theta)$ with the gap function given
by the vortex form
\begin{equation}
 \Delta(x,y) = |\Delta(r)|e^{in\theta}
  \quad\text{with}\quad |\Delta(\infty)|>0,
\end{equation}
it is easy to find the explicit zero-energy solution for odd
$N_w=n$ (Ref.~\onlinecite{Herbut:2010})
\begin{equation}\label{eq:solution_rela_2D}
 \begin{split}
  \begin{pmatrix}
   u_1 \\ u_2
  \end{pmatrix}
  &=
  \begin{bmatrix}
   \sqrt{\mu+h}\,J_{l}\!\left(\sqrt{\mu^2-h^2}\,r\right)e^{-\frac\pi4i} \\
   \sqrt{\mu-h}\,J_{l+1}\!\left(\sqrt{\mu^2-h^2}\,r\right)e^{\frac\pi4i+i\theta}
  \end{bmatrix} \\
  &\quad \times e^{il\theta-\int^rdr'|\Delta(r')|}
 \end{split}
\end{equation}
and $v_1=-u_1^*$, $v_2=u_2^*$ with an integer $l\equiv(n-1)/2$.  We note
that the zero-energy solution, Eq.~(\ref{eq:solution_rela_2D}), is
normalizable as long as
\begin{equation}\label{eq:condition_rela_2D}
 \mu^2+|\Delta(\infty)|^2 > h^2
\end{equation}
is satisfied and there is a topological phase transition at
$\mu^2+|\Delta|^2=h^2$ [see also Eq.~(\ref{eq:chern_rela}) below].

\subsection{Derivation of $p_x+ip_y$ superconductor and fermion zero mode}
We now derive a clear connection between the Jackiw-Rossi model and the
nonrelativistic $p_x+ip_y$ superconductor.  Suppose we are interested in
the low-energy spectrum of Hamiltonian (\ref{eq:H_rela_2D}) in the limit
where both $h>0$ and $\mu>0$ are equally large
\begin{equation}\label{eq:limit_2D}
 \varepsilon,\,|\sqrt{\mu^2+|\Delta|^2}-h| \ll h \sim \mu.
\end{equation}
The low-energy spectrum in such a limit can be obtained by eliminating
small components $u_2$ and $v_2$~\cite{Capelle:1995}.  Substituting the
following two equations from Eq.~(\ref{eq:eigenvalue_2D}):
\begin{equation}
 \begin{split}
  \left(\varepsilon+h+\mu\right)u_2 &= p_+u_1+\Delta v_1 \\
  \left(\varepsilon-h-\mu\right)v_2 &= -p_-v_1+\Delta^*u_1
 \end{split}
\end{equation}
into the remaining two equations, we obtain
\begin{equation}
 \begin{split}
  \left(\varepsilon-h+\mu\right)u_1
  &= \frac{p^2u_1+p_-\Delta v_1}{\varepsilon+h+\mu}
  + \frac{-\Delta p_-v_1+|\Delta|^2u_1}{\varepsilon-h-\mu} \\
  \left(\varepsilon+h-\mu\right)v_1
  &= \frac{p^2v_1-p_+\Delta^*u_1}{\varepsilon-h-\mu}
  + \frac{\Delta^*p_+u_1+|\Delta|^2v_1}{\varepsilon+h+\mu}.
 \end{split}
\end{equation}
Here we introduced $p_\pm\equiv p_x\pm ip_y$.

In the limit under consideration, Eq.~(\ref{eq:limit_2D}), we can
neglect $\varepsilon$ compared to $h+\mu$ and approximate
$\sqrt{\mu^2+|\Delta|^2}$ by $h$.  The remaining components $u_1$ and
$v_1$ obey the new energy eigenvalue problem
\begin{equation}\label{eq:new_eigenvalue_2D}
 \varepsilon
  \begin{pmatrix}
   u_1 \\ v_1
  \end{pmatrix}
  =
  \begin{pmatrix}
   \frac{p^2}{2m}-\mu_\nr
   & \frac12\left\{p_-,\Delta_\nr\right\} \\
   \frac12\left\{p_+,\Delta_\nr^*\right\}
   & -\frac{p^2}{2m}+\mu_\nr
  \end{pmatrix}
  \begin{pmatrix}
   u_1 \\ v_1
  \end{pmatrix},
\end{equation}
where we defined the nonrelativistic mass, chemical potential, and
pairing gap as
\begin{equation}\label{eq:definition_2D}
 m \equiv h,\qquad \mu_\nr \equiv \sqrt{\mu^2+|\Delta|^2}-h,
  \qquad \Delta_\nr \equiv \frac{\Delta}{h}.
\end{equation}
The resulting Hamiltonian
\begin{equation}\label{eq:H_nonrela_2D}
 \H_\nr =
  \begin{pmatrix}
   \frac{p^2}{2m}-\mu_\nr
   & \frac12\left\{p_-,\Delta_\nr\right\} \\
   \frac12\left\{p_+,\Delta_\nr^*\right\}
   & -\frac{p^2}{2m}+\mu_\nr
  \end{pmatrix}
\end{equation}
describes the nonrelativistic $p_x+ip_y$ superconductor.  We note that
when $h<0$, one obtains the Hamiltonian of the $p_x-ip_y$ superconductor
where $p_+$ and $p_-$ are exchanged in Eq.~(\ref{eq:H_nonrela_2D}).

The first nontrivial check of this correspondence is the comparison of
spectrum in a uniform space where $\Delta$ is constant.  The
relativistic Hamiltonian (\ref{eq:H_rela_2D}) has the energy eigenvalues
\begin{equation}
 \varepsilon^2 = p^2+h^2+\mu^2+|\Delta|^2
  \pm2\sqrt{p^2\mu^2+h^2\left(\mu^2+|\Delta|^2\right)}.
\end{equation}
Its low-energy branch (lower sign) at small $p$ is correctly reproduced
by the energy eigenvalue of the nonrelativistic Hamiltonian
(\ref{eq:H_nonrela_2D})
\begin{equation}
 \varepsilon_\nr^2 = \left(\frac{p^2}{2m}-\mu_\nr\right)^2+p^2|\Delta_\nr|^2
\end{equation}
under the assumptions in Eq.~(\ref{eq:limit_2D}).

Because the above ``nonrelativistic limit'' does not rely on the spatial
independence of $\Delta$, the fermion zero mode found in
Eq.~(\ref{eq:solution_rela_2D}) persists into the $p_x+ip_y$
superconductor, Eq.~(\ref{eq:H_nonrela_2D}).  In order to demonstrate
it, we consider the simplified vortex configuration with a constant
$|\Delta_\nr|>0$
\begin{equation}
 \Delta_\nr(x,y) = e^{in\theta}|\Delta_\nr|.
\end{equation}
When $n$ is odd, we can find the explicit zero-energy solution
($\varepsilon=0$) to
Eq.~(\ref{eq:new_eigenvalue_2D}) (Ref.~\onlinecite{Gurarie:2007})
\begin{equation}\label{eq:solution_nonrela_2D}
 u_1 = J_{l}\!\left[\sqrt{2m\mu_\nr-\left(m|\Delta_\nr|\right)^2}\,r\right]
  e^{-\frac\pi4i+il\theta-m|\Delta_\nr|r}
\end{equation}
and $v_1=-u_1^*$.  One can see that this zero-energy solution is the
direct consequence of that in Eq.~(\ref{eq:solution_rela_2D}) because
Eqs.~(\ref{eq:limit_2D}) and (\ref{eq:definition_2D}) lead to
\begin{equation}
 \mu^2-h^2 = \left(\mu+h\right)\left(\mu-h\right)
  \approx 2m\left(\mu_\nr-\frac{m|\Delta_\nr|^2}{2}\right).
\end{equation}

Thus we have established that the existence of a fermion zero mode bound
to a vortex in the $p_x+ip_y$ superconductor,
Eq.~(\ref{eq:H_nonrela_2D}), is a remnant of that in the Jackiw-Rossi
model, Eq.~(\ref{eq:H_rela_2D}).  In particular, the condition for the
normalizability of the zero-energy solution,
Eq.~(\ref{eq:condition_rela_2D}), is translated into
\begin{equation}\label{eq:condition_nonrela_2D}
 \mu_\nr > 0
\end{equation}
which coincides with the well-known topological phase transition in the
$p_x+ip_y$ superconductor existing at
$\mu_\nr=0$ (Refs.~\onlinecite{Volovik:book,Read:1999fn,Gurarie:2007})
[see also Eq.~(\ref{eq:chern_nonrela}) below].  Our finding also
clarifies why a vortex with winding number $N_w$ in the $p_x+ip_y$
superconductor cannot support $|N_w|$ zero modes in contrast to in the
Jackiw-Rossi model with $h=\mu=0$~\cite{Tewari:2007,Gurarie:2007}.  In
order to derive the $p_x+ip_y$ superconductor as a nonrelativistic limit
of the Jackiw-Rossi model, one needs to introduce $h$ and $\mu$ which
split an even number of zero modes into positive- and negative-energy
states.  Therefore, only one zero mode survives for odd $N_w$ in the
$p_x+ip_y$ superconductor.

\subsection{Altland-Zirnbauer symmetry class
(Refs.~\onlinecite{Altland:1997,Zirnbauer:1996}) and topological invariant}

\begin{table}[t]
 \caption{Properties under the time-reversal operator $\T$.
 $\tau$-matrices act on the particle-hole space and $\circ$ ($\times$)
 indicates even (odd) under $\T$.  Replacement of $\tau_0$ by $\tau_3$
 exchanges the roles of $\Delta_1$ and $\Delta_2$.  \label{tab:time_2D}}
 \begin{ruledtabular}
  \begin{tabular}{ccc|cccc}
   $\T$ & $\T^T/\T$
   && $h$ & $\mu$ & $\Delta_1$ & $\Delta_2$ \\\hline
   $\sigma_2\otimes\tau_0$ & $-1$ && $\times$ & $\circ$ & $\circ$ & $\times$ \\
   $\sigma_1\otimes\tau_1$ & $+1$ && $\times$ & $\times$ & $\circ$ & $\circ$ \\
  \end{tabular}
 \end{ruledtabular}
\end{table}

Finally, we note the Altland-Zirnbauer symmetry class of the
Hamiltonians that we have investigated in this section.  Hamiltonian
(\ref{eq:H_rela_2D}) with spatially dependent $\Delta_{1,2}\neq0$ has
the charge conjugation symmetry
\begin{equation}
 \C^{-1}\H\C = -\H^* \qquad\text{with}\qquad \C =
  \begin{pmatrix}
   0 & -i\sigma_2 \\ i\sigma_2 & 0
  \end{pmatrix}.
\end{equation}
The properties of each term under the time-reversal operator
$\T$ ($\T^{-1}\H\T=\H^*$ at $h=\mu=\Delta_{1,2}=0$) are summarized in
Table~\ref{tab:time_2D}.  In particular, the so-called chiral symmetry,
\begin{equation}\label{eq:chiral_2D}
 \chi^{-1}\H\chi = -\H \qquad\text{with}\qquad \chi =
  \begin{pmatrix}
   \sigma_3 & 0 \\ 0 & -\sigma_3
  \end{pmatrix},
\end{equation}
is present only if $h=\mu=0$ and essential for the index theorem,
Eq.~(\ref{eq:index_2D}).  Therefore, the relativistic Hamiltonian
(\ref{eq:H_rela_2D}) with $h,\mu\neq0$ and thus the resulting
nonrelativistic Hamiltonian (\ref{eq:H_nonrela_2D}) belong to the
symmetry class D.

According to Refs.~\onlinecite{Schnyder:2008} and
\onlinecite{Kitaev:2009}, the class D Hamiltonians defined in compact 2D
momentum spaces can be classified by an integer-valued topological
invariant, which is the first Chern
number~\cite{Thouless:1982,Kohmoto:1985,Niu:1985}
\begin{equation}\label{eq:chern1}
 C_1 \equiv \frac{-i}{2\pi}\int\!d\p 
  \left(\frac{\d a_y}{\d p_x}-\frac{\d a_x}{\d p_y}\right)
\end{equation}
with
\begin{equation}\label{eq:chern2}
 a_i(\p) \equiv \sum_{\varepsilon_a<0}\left\langle\varepsilon_a,\p\right|
  \frac{\d}{\d p_i}\left|\varepsilon_a,\p\right\rangle.
\end{equation}
We shall use Eqs.~(\ref{eq:chern1}) and (\ref{eq:chern2}) as a
definition of the topological invariant $C_1$ even for relativistic
(Dirac-type) Hamiltonians while $C_1$ in this case can be a
half integer.  However, for superconductors, $C_1$ is always an integer
because of the Nambu-Gor'kov doubling.  We find that the topological
invariant for the relativistic Hamiltonian (\ref{eq:H_rela_2D}) is given
by
\begin{equation}\label{eq:chern_rela}
 C_1 =
  \begin{cases}
   \quad\ \,0  &\ \text{for}\ \ \ \mu^2+|\Delta|^2>h^2 \\
   -\mathrm{sgn}(h) &\ \text{for}\ \ \ \mu^2+|\Delta|^2<h^2,
  \end{cases}
\end{equation}
while the topological invariant for the nonrelativistic Hamiltonian
(\ref{eq:H_nonrela_2D}) is given by
\begin{equation}\label{eq:chern_nonrela}
 C_1 =
  \begin{cases}
   \,1 &\ \text{for}\ \ \ \mu_\nr>0 \\
   \,0 &\ \text{for}\ \ \ \mu_\nr<0.
  \end{cases}
\end{equation}
Therefore, in general, the topological invariant of the momentum space
Hamiltonian is not preserved by the nonrelativistic limit.\footnote{The
  topological invariant can be matched if we properly regularize the
  large $\p$ behavior of the relativistic Hamiltonian: Replacing $h$ in
  Eq.~(\ref{eq:H_rela_2D}) by $h\left(1+|\epsilon|\p^2\right)$, the
  Chern number becomes $\mathrm{sgn}(h)$ for $\mu^2+|\Delta|^2>h^2$ and
  $0$ for $\mu^2+|\Delta|^2<h^2$, which coincides with that of
  Eq.~(\ref{eq:H_nonrela_2D}).}

Nevertheless, both values of $C_1$ computed for the relativistic and
nonrelativistic Hamiltonians are consistent with recent conjectures
relating the topological invariant of a momentum space Hamiltonian to
the number of fermion zero modes bound to a
vortex~\cite{Teo:2010zb,Santos:2009}.  For class D superconductors
defined in compact momentum spaces (as is the case for nonrelativistic
Hamiltonians), Teo and Kane in Ref.~\onlinecite{Teo:2010zb} conjecture
that the number of fermion zero modes is
\begin{equation}
 \nu = C_1N_w \quad \mathrm{mod}\ 2.
\end{equation}
This formula gives $\nu=1$ for $\mu_\nr>0$ and $\nu=0$ for $\mu_\nr<0$
for an odd winding number $N_w$.  On the other hand, Santos
{\it et al.}\ in Ref.~\onlinecite{Santos:2009} do not constrain
Hamiltonians to be defined in compact momentum spaces, allowing for
relativistic (Dirac-type) Hamiltonians, and conjecture that the number
of fermion zero modes is
\begin{equation}
 \nu = \left(C_1+N_f\right)N_w \quad \mathrm{mod}\ 2,
\end{equation}
where $N_f$ is the number of Dirac flavors [$N_f=1$ for the
Jackiw-Rossi model, Eq.~(\ref{eq:H_rela_2D}), and $N_f=0$ for the
$p_x+ip_y$ superconductor, Eq.~(\ref{eq:H_nonrela_2D})].  For an odd
winding number $N_w$, their formula gives $\nu=1$ for
$\mu^2+|\Delta|^2>h^2$ and $\mu_\nr>0$ and $\nu=0$ for
$\mu^2+|\Delta|^2<h^2$ and $\mu_\nr<0$.  Therefore, the conjectured
counting of fermion zero modes in terms of the momentum space
topological invariant works both in the relativistic and nonrelativistic
Hamiltonians, even though the value of $C_1$ is not preserved by the
nonrelativistic limit.

\section{Jackiw-Rebbi model in 3D and~its~nonrelativistic~limit \label{sec:3D}}

\subsection{Jackiw-Rebbi model and fermion zero mode at a hedgehog}
In this section, we extend the above developed analysis to three
dimensions.  For this purpose, we consider the following Hamiltonian
describing 3D Dirac fermions coupled with three real scalar fields
($\Delta\equiv\Delta_1+i\Delta_2$ and $\Delta_3$):
\begin{equation}
 H = \frac12\int\!d\x\,\Psi^\+\H\Psi
\end{equation}
with $\Psi^\+=(\psi^\+,-i\psi^T\alpha_2)$ and
\begin{equation}\label{eq:H_rela_3D}
 \begin{split}
  & \H = \\ &
  \begin{pmatrix}
   \bm\alpha\cdot\p+\beta m-\mu-i\gamma^5\beta\Delta_3 & \Delta \\
   \Delta^* & -\bm\alpha\cdot\p+\beta m+\mu+i\gamma^5\beta\Delta_3
  \end{pmatrix}.
 \end{split}
\end{equation}
This Hamiltonian with zero mass $m$ and zero chemical potential $\mu$,
after an appropriate unitary transformation and renamings
($\beta\leftrightarrow i\gamma^5\beta,\Delta_1\to\phi_1,\Delta_2\to-\phi_2,\Delta_3\to-\phi_3$),
was studied initially by Jackiw and Rebbi~\cite{Jackiw:1975fn} and
recently by Teo and Kane~\cite{Teo:2009qv} in the context of ordinary
and topological insulators coexisting with superconductivity.  When the
scalar fields $\Delta_{1,2,3}$ are spatially dependent,
$\p\equiv(p_x,p_y,p_z)$ has to be regarded as derivative operators
$(-i\d_x,-i\d_y,-i\d_z)$. The energy eigenvalue problem is
\begin{equation}\label{eq:eigenvalue_3D}
 \varepsilon
  \begin{pmatrix}
   u_1 \\ u_2 \\ v_2 \\ v_1
  \end{pmatrix}
  = \H
  \begin{pmatrix}
   u_1 \\ u_2 \\ v_2 \\ v_1
  \end{pmatrix},
\end{equation}
where $u_{1,2}$ and $v_{1,2}$ are two-component fields.  Here we employ
the standard representation of Dirac matrices
\begin{equation}
 \bm\alpha =
  \begin{pmatrix}
   0 & \bm\sigma \\ \bm\sigma & 0
  \end{pmatrix},\quad 
  \beta =
  \begin{pmatrix}
   \openone & 0 \\ 0 & -\openone
  \end{pmatrix},\quad
  \gamma^5 =
  \begin{pmatrix}
   0 & \openone \\ \openone & 0
  \end{pmatrix}
\end{equation}
and hence
\begin{equation}
 i\gamma^5\beta =
  \begin{pmatrix}
   0 & -i\openone \\ i\openone & 0
  \end{pmatrix}.
\end{equation}

When $m$ and $\mu$ are both zero, the number of fermion zero modes
($\varepsilon=0$) bound to a hedgehog formed by $\Delta_{1,2,3}(x,y,z)$
is determined by the winding number of the three scalar
fields~\cite{Jackiw:1975fn,Callias:1977kg,Fukui:2010a}
\begin{equation}\label{eq:index_3D}
 \mathrm{Index}\,\H = \frac1{8\pi}\int\!dS_i\,\epsilon_{ijk}
  \epsilon_{abc}\hat\Delta_a\d_j\hat\Delta_b\d_k\hat\Delta_c \equiv N_w,
\end{equation}
where
$\hat\Delta_a\equiv\Delta_a/\sqrt{\Delta_1^2+\Delta_2^2+\Delta_3^2}$ and
the surface integral is taken at spatial infinity.  However, in the
presence of $m$ and $\mu$, the index theorem is no longer valid:  $m$
and $\mu$ terms in the Hamiltonian can couple zero modes and they become
nonzero energy states so that two states form a pair with opposite
energies.  Therefore, in general, only one zero mode survives for odd
$N_w$ while no zero mode survives for even
$N_w$~\cite{Fukui:2010a,Teo:2010zb}.\footnote{Two exceptional cases are
  $m=\pm\mu$.  When $N_w>0\,(<0)$, the zero-energy solutions at
  $m=\mu=0$ still solve Eq.~(\ref{eq:eigenvalue_3D}) with
  $m=+(-)\mu\neq0$ and thus there are $|N_w|$ zero modes.  This can be
  easily seen if one rewrites the Hamiltonian in the basis where the
  chiral operator defined in Eq.~(\ref{eq:chiral_3D}) has the form
  $\chi=\mathrm{diag}(\openone,\openone,-\openone,-\openone)$ and
  recognizes that the $|N_w|$ zero-energy solutions are eigenstates of
  $\chi$ with the eigenvalue $+1\,(-1)$ for $N_w>0\,(<0)$.}

Here, instead of the symmetric hedgehog ($\Delta_i\propto\hat{x}_i$), we
assume the hedgehoglike configuration in which $\Delta_{1,2}$ depend
only on $(x,y)$ and form a vortex and $\Delta_3$ depends only on $z$ and
forms a kink.  They have the same winding number but the latter has the
advantage that an analytic solution can be found even with $m,\mu\neq0$.
If we work in cylindrical coordinates $(r,\theta,z)$ with the gap
functions given by the forms
\begin{equation}\label{eq:vortex_3D}
 \Delta(x,y) = |\Delta(r)|e^{in\theta}
  \quad\text{with}\quad |\Delta(\infty)|>0
\end{equation}
and
\begin{equation}\label{eq:kink_3D-1}
 \Delta_3(z\to\pm\infty) \to \pm|\Delta_3|,
\end{equation}
it is easy to find the explicit zero-energy solution for odd
$N_w=n$ (Ref.~\onlinecite{Fukui:2010b})
\begin{equation}\label{eq:solution_rela_3D-1}
 \begin{split}
  \begin{pmatrix}
   u_1 \\ u_2
  \end{pmatrix}
  &=
  \begin{bmatrix}
   \sqrt{\mu+m}\,J_{l}\!\left(\sqrt{\mu^2-m^2}\,r\right)e^{-\frac\pi4i} \\
   0 \\
   0 \\
   \sqrt{\mu-m}\,J_{l+1}\!\left(\sqrt{\mu^2-m^2}\,r\right)e^{\frac\pi4i+i\theta}
  \end{bmatrix} \\
  &\quad \times e^{il\theta-\int^rdr'|\Delta(r')|-\int^zdz'\Delta_3(z')}
 \end{split}
\end{equation}
and $v_1=i\sigma_2u_1$, $v_2=i\sigma_2u_2$ with an integer
$l\equiv(n-1)/2$.  On the other hand, when
\begin{equation}\label{eq:kink_3D-2}
 \Delta_3(z\to\pm\infty) \to \mp|\Delta_3|
\end{equation}
with the same $\Delta(x,y)$ in Eq.~(\ref{eq:vortex_3D}), we have $N_w=-n$
and the zero-energy solution in Eq.~(\ref{eq:solution_rela_3D-1}) is
replaced by
\begin{equation}\label{eq:solution_rela_3D-2}
 \begin{split}
  \begin{pmatrix}
   u_1 \\ u_2
  \end{pmatrix}
  &=
  \begin{bmatrix}
   0 \\
   \sqrt{\mu+m}\,J_{l+1}\!\left(\sqrt{\mu^2-m^2}\,r\right)e^{\frac\pi4i+i\theta} \\
   \sqrt{\mu-m}\,J_{l}\!\left(\sqrt{\mu^2-m^2}\,r\right)e^{-\frac\pi4i} \\
   0
  \end{bmatrix} \\
  &\quad \times e^{il\theta-\int^rdr'|\Delta(r')|+\int^zdz'\Delta_3(z')}.
 \end{split}
\end{equation}
We note that the zero-energy solution, Eq.~(\ref{eq:solution_rela_3D-1})
or (\ref{eq:solution_rela_3D-2}), is normalizable as long as
\begin{equation}\label{eq:condition_rela_3D}
 \mu^2+|\Delta(\infty)|^2 > m^2
\end{equation}
is satisfied.

\subsection{Derivation of $p+is$ superconductor and fermion zero mode}
We now study the nonrelativistic limit of the above Jackiw-Rebbi model.
Suppose we are interested in the low-energy spectrum of Hamiltonian
(\ref{eq:H_rela_3D}) in the limit where both $m>0$ and $\mu>0$ are
equally large
\begin{equation}\label{eq:limit_3D}
 \varepsilon,\,|\sqrt{\mu^2+|\Delta|^2}-m| \ll m \sim \mu.
\end{equation}
The low-energy spectrum in such a limit can be obtained by eliminating
small components $u_2$ and $v_2$~\cite{Capelle:1995}.  Substituting the
following two equations from Eq.~(\ref{eq:eigenvalue_3D}):
\begin{equation}
 \begin{split}
  \left(\varepsilon+m+\mu\right)u_2 
  &= \left(\bm\sigma\cdot\p-i\Delta_3\right)u_1+\Delta v_1 \\
  \left(\varepsilon-m-\mu\right)v_2
  &= -\left(\bm\sigma\cdot\p+i\Delta_3\right)v_1+\Delta^*u_1
 \end{split}
\end{equation}
into the remaining two equations, we obtain
\begin{equation}
 \begin{split}
  & \left(\varepsilon-m+\mu\right)u_1 \\
  &\quad = \frac{\left[p^2+\Delta_3^2-\bm\sigma\cdot\left(\bm\d\Delta_3\right)\right]u_1
  +\left(\bm\sigma\cdot\p+i\Delta_3\right)\Delta v_1}{\varepsilon+m+\mu} \\
  &\qquad +\frac{-\Delta\left(\bm\sigma\cdot\p+i\Delta_3\right)v_1
  +|\Delta|^2u_1}{\varepsilon-m-\mu} \\
  & \left(\varepsilon+m-\mu\right)v_1 \\
  &\quad = \frac{\left[p^2+\Delta_3^2+\bm\sigma\cdot\left(\bm\d\Delta_3\right)\right]v_1
  -\left(\bm\sigma\cdot\p-i\Delta_3\right)\Delta^*u_1}{\varepsilon-m-\mu} \\
  &\qquad +\frac{\Delta^*\left(\bm\sigma\cdot\p-i\Delta_3\right)u_1
  +|\Delta|^2v_1}{\varepsilon+m+\mu}.
 \end{split}
\end{equation}
Here the derivative operator $\bm\d$ in
$\bm\sigma\cdot\left(\bm\d\Delta_3\right)$ acts only on $\Delta_3$.

In the limit under consideration, Eq.~(\ref{eq:limit_3D}), we can
neglect $\varepsilon$ compared to $m+\mu$ and approximate
$\sqrt{\mu^2+|\Delta|^2}$ by $m$.  The remaining components $u_1$ and
$v_1$ obey the new energy eigenvalue problem
\begin{equation}\label{eq:new_eigenvalue_3D}
 \begin{split}
  & \varepsilon
  \begin{pmatrix}
   u_1 \\ v_1
  \end{pmatrix}
  = \\ &
  \begin{bmatrix}
   \frac{p^2}{2m}-\mu_\nr-\frac{\bm\sigma\cdot\left(\bm\d\Delta_3\right)}{2m}
   & \frac12\left\{\bm\sigma\cdot\p,\Delta_t\right\}+i\Delta_s \\
   \frac12\left\{\bm\sigma\cdot\p,\Delta_t^*\right\}-i\Delta_s^*
   & -\frac{p^2}{2m}+\mu_\nr-\frac{\bm\sigma\cdot\left(\bm\d\Delta_3\right)}{2m}
  \end{bmatrix}
  \begin{pmatrix}
   u_1 \\ v_1
  \end{pmatrix},
 \end{split}
\end{equation}
where we defined the nonrelativistic chemical potential as
\begin{equation}\label{eq:definition_3D-1}
 \mu_\nr \equiv \sqrt{\mu^2+|\Delta|^2}-m-\frac{\Delta_3^2}{2m}
\end{equation}
and the spin-triplet $p$-wave and spin-singlet $s$-wave pairing gaps as
\begin{equation}\label{eq:definition_3D-2}
 \Delta_t \equiv \frac{\Delta}{m} \qquad\text{and}\qquad
  \Delta_s \equiv \frac{\Delta_3\Delta}{m}.
\end{equation}
The resulting Hamiltonian
\begin{equation}\label{eq:H_nonrela_3D}
 \H_\nr =
  \begin{bmatrix}
   \frac{p^2}{2m}-\mu_\nr-\frac{\bm\sigma\cdot\left(\bm\d\Delta_3\right)}{2m}
   & \frac12\left\{\bm\sigma\cdot\p,\Delta_t\right\}+i\Delta_s \\
   \frac12\left\{\bm\sigma\cdot\p,\Delta_t^*\right\}-i\Delta_s^*
   & -\frac{p^2}{2m}+\mu_\nr-\frac{\bm\sigma\cdot\left(\bm\d\Delta_3\right)}{2m}
  \end{bmatrix}
\end{equation}
describes the $p+is$ superconductor in which spin-triplet $p$-wave and
spin-singlet $s$-wave pairings coexist.  $\Delta_s$ can be complex but
its phase is locked to the phase of $\Delta_t$ [see
Eq.~(\ref{eq:definition_3D-2})] and thus there are three independent
degrees of freedom.  The last term in the diagonal elements resembles
the Zeeman coupling $\bm\sigma\cdot\bm{B}$ with ``magnetic field''
$B_i=-\d_i\Delta_3/(2m)$ generated by the gradient of
$\Delta_3=\Delta_s/\Delta_t$.  We note that the nonrelativistic
Hamiltonian (\ref{eq:H_nonrela_3D}) in the absence of $\Delta_s$ is the
BW state of the superfluid $^3$He and studied in
Ref.~\onlinecite{Silaev:2010}.

The first nontrivial check of this correspondence is the comparison of
spectrum in a uniform space where $\Delta$ and $\Delta_3$ are constant.
The relativistic Hamiltonian (\ref{eq:H_rela_3D}) has the energy
eigenvalues
\begin{equation}
 \begin{split}
  \varepsilon^2 &= p^2+m^2+\mu^2+|\Delta|^2+\Delta_3^2 \\
  &\quad \pm2\sqrt{p^2\mu^2+m^2\left(\mu^2+|\Delta|^2\right)+\mu^2\Delta_3^2}.
 \end{split}
\end{equation}
Its low-energy branch (lower sign) at small $p$ and $\Delta_3$ is
correctly reproduced by the energy eigenvalue of the nonrelativistic
Hamiltonian (\ref{eq:H_nonrela_3D})
\begin{equation}
 \varepsilon_\nr^2 = \left(\frac{p^2}{2m}-\mu_\nr\right)^2
  +p^2|\Delta_t|^2+|\Delta_s|^2
\end{equation}
under the assumptions in Eq.~(\ref{eq:limit_3D}).

Because the above nonrelativistic limit does not rely on the spatial
independence of $\Delta$ and $\Delta_3$, the fermion zero mode found in
Eq.~(\ref{eq:solution_rela_3D-1}) or (\ref{eq:solution_rela_3D-2})
persists into the $p+is$ superconductor, Eq.~(\ref{eq:H_nonrela_3D}).
In order to demonstrate it, we consider the simplified hedgehoglike
configuration resulting from Eqs.~(\ref{eq:vortex_3D}),
(\ref{eq:kink_3D-1}), and (\ref{eq:kink_3D-2}) with constant
$|\Delta_t|>0$ and $|\Delta_s|>0$:
\begin{equation}\label{eq:hedgehog_3D}
 \begin{split}
  & \Delta_t(x,y) = e^{in\theta}|\Delta_t| \quad\text{and} \\
  & \Delta_s(x,y,z) = \pm e^{in\theta}\mathrm{sgn}(z)|\Delta_s|.
 \end{split}
\end{equation}
When $n$ is odd, we can find the explicit zero-energy solution
($\varepsilon=0$) to Eq.~(\ref{eq:new_eigenvalue_3D})
\begin{equation}\label{eq:solution_nonrela_3D-1}
 \begin{split}
  u_1 &=
  \begin{Bmatrix}
   J_{l}\!\left[\sqrt{2m\mu_\nr-\left(m|\Delta_t|\right)^2
   +\left|\frac{\Delta_s}{\Delta_t}\right|^2}\,r\right] \\
   0
  \end{Bmatrix} \\
  &\quad \times e^{-\frac\pi4i+il\theta
  -m|\Delta_t|r-\left|\frac{\Delta_s}{\Delta_t}\right||z|}
 \end{split}
\end{equation}
corresponding to the upper sign in Eq.~(\ref{eq:hedgehog_3D}), or
\begin{equation}\label{eq:solution_nonrela_3D-2}
 \begin{split}
  u_1 &=
  \begin{Bmatrix}
   0 \\
   J_{l+1}\!\left[\sqrt{2m\mu_\nr-\left(m|\Delta_t|\right)^2
   +\left|\frac{\Delta_s}{\Delta_t}\right|^2}\,r\right]
  \end{Bmatrix} \\
  &\quad \times e^{\frac\pi4i+i\left(l+1\right)\theta
  -m|\Delta_t|r-\left|\frac{\Delta_s}{\Delta_t}\right||z|}
 \end{split}
\end{equation}
corresponding to the lower sign in Eq.~(\ref{eq:hedgehog_3D}), and
$v_1=i\sigma_2u_1^*$.  One can see that this zero-energy solution is the
direct consequence of that in Eq.~(\ref{eq:solution_rela_3D-1}) or
(\ref{eq:solution_rela_3D-2}) because Eqs.~(\ref{eq:limit_3D}),
(\ref{eq:definition_3D-1}), and (\ref{eq:definition_3D-2}) lead to
\begin{equation}
 \mu^2-m^2 
  \approx 2m\left[\mu_\nr-\frac{m|\Delta_t|^2}{2}
	     +\frac{1}{2m}\left(\frac{\Delta_s}{\Delta_t}\right)^2\right].
\end{equation}

Thus we have established that the existence of a fermion zero mode bound
to the hedgehoglike structure, Eq.~(\ref{eq:hedgehog_3D}), formed by
$\Delta_t$ and $\Delta_s/\Delta_t$ in the $p+is$ superconductor,
Eq.~(\ref{eq:H_nonrela_3D}), is a remnant of that in the Jackiw-Rebbi
model, Eq.~(\ref{eq:H_rela_3D}).  In particular, the condition for the
normalizability of the zero-energy solution,
Eq.~(\ref{eq:condition_rela_3D}), is translated into
\begin{equation}\label{eq:condition_nonrela_3D}
 \mu_\nr+\frac{1}{2m}\left(\frac{\Delta_s}{\Delta_t}\right)^2 > 0.
\end{equation}

\subsection{Altland-Zirnbauer symmetry class
(Refs.~\onlinecite{Altland:1997,Zirnbauer:1996})}

\begin{table}[t]
 \caption{Properties under the time-reversal operator $\T$.
 $\tau$-matrices act on the particle-hole space and $\circ$ ($\times$)
 indicates even (odd) under $\T$.  Replacement of $\tau_0$ by $\tau_3$
 exchanges the roles of $\Delta_1$ and $\Delta_2$.  \label{tab:time_3D}}
 \begin{ruledtabular}
  \begin{tabular}{ccc|ccccc}
   $\T$ & $\T^T/\T$
   && $\mu$ & $m$ & $\Delta_3$ & $\Delta_1$ & $\Delta_2$ \\\hline
   $\phantom{\gamma^5}\alpha_2\otimes\tau_0$ & $-1$
       && $\circ$ & $\times$ & $\circ$ & $\circ$ & $\times$ \\
   $\gamma^5\alpha_2\otimes\tau_0$ & $-1$
       && $\circ$ & $\circ$ & $\times$ & $\circ$ & $\times$ \\
   $\phantom{\gamma^5}\beta\,\alpha_2\otimes\tau_1$ & $+1$
       && $\times$ & $\times$ & $\circ$ & $\circ$ & $\circ$ \\
   $\gamma^5\beta\,\alpha_2\otimes\tau_1$ & $-1$ 
       && $\times$ & $\circ$ & $\times$ & $\circ$ & $\circ$ \\
  \end{tabular}
 \end{ruledtabular}
\end{table}

Finally, we note the Altland-Zirnbauer symmetry class of the
Hamiltonians that we have investigated in this section.  Hamiltonian
(\ref{eq:H_rela_3D}) with spatially dependent $\Delta_{1,2,3}\neq0$ has
the charge conjugation symmetry
\begin{equation}
 \C^{-1}\H\C = -\H^* \qquad\text{with}\qquad \C =
  \begin{pmatrix}
   0 & -i\alpha_2 \\ i\alpha_2 & 0
  \end{pmatrix}.
\end{equation}
The properties of each term under the time-reversal operator
$\T$ ($\T^{-1}\H\T=\H^*$ at $m=\mu=\Delta_{1,2,3}=0$) are summarized in
Table~\ref{tab:time_3D}.  In particular, the so-called chiral symmetry,
\begin{equation}\label{eq:chiral_3D}
 \chi^{-1}\H\chi = -\H \qquad\text{with}\qquad \chi =
  \begin{pmatrix}
   \beta & 0 \\ 0 & -\beta
  \end{pmatrix},
\end{equation}
is present only if $m=\mu=0$ and essential for the index theorem,
Eq.~(\ref{eq:index_3D}).  Therefore, the relativistic Hamiltonian
(\ref{eq:H_rela_3D}) with $m,\mu\neq0$ and thus the resulting
nonrelativistic Hamiltonian (\ref{eq:H_nonrela_3D}) belong to the
symmetry class D.  There is no topological classification of class D
Hamiltonians in 3D momentum spaces~\cite{Schnyder:2008,Kitaev:2009}.

\section{Summary \label{sec:summary}}
We have studied the nonrelativistic limit of the Jackiw-Rossi model in
2D and the Jackiw-Rebbi model in 3D, both of which are known to
exhibit fermion zero modes associated with pointlike topological defects
(vortex and hedgehog).  We showed that the nonrelativistic limit of the
2D Jackiw-Rossi model leads to the $p_x+ip_y$ superconductor.  Because
the fermion zero mode persists under taking this limit, we obtain a
clear understanding of the existence of a fermion zero mode bound to a
vortex in the $p_x+ip_y$ superconductor as a remnant of that in the
Jackiw-Rossi model.  Similarly, the nonrelativistic limit of the 3D
Jackiw-Rebbi model leads to the $p+is$ superconductor in which the
spin-triplet $p$-wave pairing gap $\Delta_t$ and the spin-singlet
$s$-wave pairing gap $\Delta_s$ coexist.  We showed that the resulting
Hamiltonian supports a fermion zero mode when $\Delta_t$ and
$\Delta_s/\Delta_t$ form a hedgehoglike structure.  Fermion zero modes
in the superconductors studied in this paper correspond to Majorana
fermions and the associated pointlike defects obey non-Abelian
statistics both in 2D (Refs.~\onlinecite{Read:1999fn,Ivanov:2001}) and
3D~\cite{Teo:2009qv,Freedman:2010ak}.

Our findings provide a unified view of Majorana zero modes in
relativistic (Dirac-type) and nonrelativistic (Schr\"odinger-type)
superconductors.  It should be possible to generalize our analysis to
other interesting cases and find new examples of nonrelativistic
Hamiltonians, which are more common in condensed matter systems, with 
topological properties that descend from Dirac-type Hamiltonians, which
are generally easier to analyze.

\acknowledgments
The authors thank M.~A.~Silaev and G.~E.~Volovik for sharing their note
regarding Ref.~\onlinecite{Silaev:2010} and R.~Jackiw and S.-Y.~Pi for
valuable discussions.  This research was supported by MIT Pappalardo
Fellowship in Physics and DOE under Grants No.\ DE-FG02-94ER40818
(Y.\,N.) and No.\ DEF-06ER46316 (C.\,C., L.\,S.).


\begin{thebibliography}{99}

\bibitem{Jackiw:1975fn}
  R.~Jackiw and C.~Rebbi,
  Phys.\ Rev.\ D {\bf 13}, 3398 (1976).

\bibitem{Teo:2010zb}
  J.~C.~Y.~Teo and C.~L.~Kane,
  Phys.\ Rev.\ B {\bf 82}, 115120 (2010).

\bibitem{Read:1999fn}
  N.~Read and D.~Green,
  Phys.\ Rev.\ B {\bf 61}, 10267 (2000).

\bibitem{Kitaev:1997wr}
  A.~Y.~Kitaev,
  Ann.\ Phys.\ {\bf 303}, 2 (2003).

\bibitem{Mackenzie:2003}
  A.~P.~Mackenzie and Y.~Maeno,
  Rev.\ Mod.\ Phys.\ {\bf 75}, 657 (2003).

\bibitem{Volovik:1999eh}
  G.~E.~Volovik,
  JETP Lett.\ {\bf 70}, 609 (1999).

\bibitem{Ivanov:2001}
  D.~A.~Ivanov,
  Phys.\ Rev.\ Lett.\ {\bf 86}, 268 (2001).

\bibitem{Jackiw:1981ee}
  R.~Jackiw and P.~Rossi,
  Nucl.\ Phys.\ B {\bf 190}, 681 (1981).

\bibitem{Fu:2008}
  L.~Fu and C.~L.~Kane,
  Phys.\ Rev.\ Lett.\ {\bf 100}, 096407 (2008).

\bibitem{Sau:2010a}
  J.~D.~Sau, R.~M.~Lutchyn, S.~Tewari, and S.~Das~Sarma,
  Phys.\ Rev.\ Lett.\ {\bf 104}, 040502 (2010).

\bibitem{Alicea:2010}
  J.~Alicea,
  Phys.\ Rev.\ B {\bf 81}, 125318 (2010).

\bibitem{Sau:2010b}
  J.~D.~Sau, S.~Tewari, R.~Lutchyn, T.~Stanescu, and S.~Das~Sarma,
  arXiv:1006.2829 [cond-mat.supr-con].

\bibitem{Lee:2009}
  P.~A.~Lee,
  arXiv:0907.2681 [cond-mat.str-el].

\bibitem{Qi:2010}
  X.-L.~Qi, T.~L.~Hughes, and S.-C.~Zhang,
  arXiv:1003.5448 [cond-mat.mes-hall].

\bibitem{Weinberg:1981eu}
  E.~J.~Weinberg,
  Phys.\ Rev.\ D {\bf 24}, 2669 (1981).

\bibitem{Tewari:2007}
  S.~Tewari, S.~Das~Sarma, and D.-H.~Lee,
  Phys.\ Rev.\ Lett.\ {\bf 99}, 037001 (2007).

\bibitem{Gurarie:2007}
  V.~Gurarie and L.~Radzihovsky,
  Phys.\ Rev.\ B {\bf 75}, 212509 (2007);
%
  see also Sec.~VIII in
  Ann.\ Phys.\ {\bf 322}, 2 (2007).

\bibitem{Fukui:2010a}
  T.~Fukui and T.~Fujiwara,
  J.\ Phys.\ Soc.\ Jpn.\ {\bf 79}, 033701 (2010).

\bibitem{Cheng:2010}
  M.~Cheng, R.~M.~Lutchyn, V.~Galitski, and S.~Das~Sarma,
  Phys.\ Rev.\ B {\bf 82}, 094504 (2010).

\bibitem{Callias:1977kg}
  C.~Callias,
  Commun.\ Math.\ Phys.\ {\bf 62}, 213 (1978).

\bibitem{Teo:2009qv}
  J.~C.~Y.~Teo and C.~L.~Kane,
  Phys.\ Rev.\ Lett.\ {\bf 104}, 046401 (2010).

\bibitem{Silaev:2010}
  M.~A.~Silaev and G.~E.~Volovik,
  arXiv:1005.4672 [cond-mat.other].

\bibitem{Nishida:2010wr}
  Y.~Nishida,
  Phys.\ Rev.\ D {\bf 81}, 074004 (2010).

\bibitem{Herbut:2010}
  The same solution with $n=1$ was obtained independently in
  I.~F.~Herbut and C.-K.~Lu,
  Phys.\ Rev.\ B {\bf 82}, 125402 (2010).

\bibitem{Capelle:1995}
  K.~Capelle and E.~K.~U.~Gross,
  Phys.\ Lett.\ A {\bf 198}, 261 (1995);
%
  Phys.\ Rev.\ B {\bf 59}, 7140 (1999);
%
  Phys.\ Rev.\ B {\bf 59}, 7155 (1999).

\bibitem{Volovik:book}
  G.~E.~Volovik,
  {\it Exotic Properties of Superfluid $^3$He\/}
  (World Scientific, Singapore, 1992);
  {\it The Universe in a Helium Droplet\/}
  (Clarendon Press, Oxford, 2003).

\bibitem{Zirnbauer:1996}
  M.~R.~Zirnbauer,
  J.\ Math.\ Phys.\ {\bf 37}, 4986 (1996).

\bibitem{Altland:1997}
  A.~Altland and M.~R.~Zirnbauer,
  Phys.\ Rev.\ B {\bf 55}, 1142 (1997).

\bibitem{Schnyder:2008}
  A.~P.~Schnyder, S.~Ryu, A.~Furusaki, and A.~W.~W.~Ludwig,
  Phys.\ Rev.\ B {\bf 78}, 195125 (2008).

\bibitem{Kitaev:2009}
  A.~Kitaev,
  AIP Conf.\ Proc.\ {\bf 1134}, 22 (2009).

\bibitem{Thouless:1982}
  D.~J.~Thouless, M.~Kohmoto, M.~P.~Nightingale, and M.~den~Nijs,
  Phys.\ Rev.\ Lett.\ {\bf 49}, 405 (1982).

\bibitem{Kohmoto:1985}
  M.~Kohmoto,
  Ann.\ Phys.\ (N.Y.) {\bf 160}, 343 (1985).

\bibitem{Niu:1985}
  Q.~Niu, D.~J.~Thouless, and Y.-S.~Wu,
  Phys.\ Rev.\ B {\bf 31}, 3372 (1985).

\bibitem{Santos:2009}
  L.~Santos, S.~Ryu, C.~Chamon, and C.~Mudry,
  Phys.\ Rev.\ B {\bf 82}, 165101 (2010).

\bibitem{Fukui:2010b}
  The same problem with $m=0$ was studied in
  T.~Fukui,
  Phys.\ Rev.\ B {\bf 81}, 214516 (2010).

\bibitem{Freedman:2010ak}
  M.~Freedman, M.~B.~Hastings, C.~Nayak, X.-L.~Qi, K.~Walker, and Z.~Wang,
  arXiv:1005.0583 [cond-mat.mes-hall].

\end{thebibliography}
\end{document}